# Integrating Artificial Intelligence and Mixed Integer Linear Programming: Explainable Graph-Based Instance Space Analysis in Air Transportation

Artur Guerra Rosa[1*], Felipe Tavares Loureiro[1], Marcus Vinicius Santos da Silva[2], Andréia Elizabeth Silva Barros[1], Silvia Araújo dos Reis[1], Victor Rafael Rezende Celestino[1]

[1]Universidade de Brasília – Brasília/DF – Brazil. E-mails: arturguerra921@hotmail.com; fefetalo@gmail.com; marcusviny63@gmail.com; andreiabarros@gmail.com; silviareis@unb.br; vrcelestino@unb.br.

[2]Pontifícia Universidade Católica de Goiás – Goiânia/GO – Brazil.

[*]Correspondence author: arturguerra921@hotmail.com

Abstract

This paper analyzes the integration of artificial intelligence (AI) with mixed integer linear programming (MILP) to address complex optimization challenges in air transportation with explainability. The study aims to validate the use of Graph Neural Networks (GNNs) for extracting structural feature embeddings from MILP instances, using the air05 crew scheduling problem. The MILP instance was transformed into a heterogeneous bipartite graph to model relationships between variables and constraints. Two neural architectures, Graph Convolutional Networks (GCN) and Graph Attention Networks (GAT) were trained to generate node embeddings. These representations were evaluated using Instance Space Analysis (ISA) through linear (PCA) and non-linear (UMAP, t-SNE) dimensionality reduction techniques. Analysis revealed that PCA failed to distinguish cluster structures, necessitating non-linear reductions to visualize the embedding topology. The GCN architecture demonstrated superior performance, capturing global topology with well-defined clusters for both variables and constraints. In contrast, the GAT model failed to organize the constraint space. The findings confirm that simpler graph architectures can effectively map the sparse topology of aviation logistics problems without manual feature engineering, contributing to explainability of instance complexity. This structural awareness provides a validated foundation for developing future "Learning to Optimize" (L2O) agents capable of improving solver performance in safety-critical environments.

## Introduction

Integrating artificial intelligence (AI) with mixed integer linear programming (MILP) techniques has become known as a transformative approach for tackling complex optimization problems across various domains, particularly in air transportation and mobility (Mirindi 2024). MILP provides a robust framework for modeling combinatorial optimization problems through linear constraints and integer variables, offering extensibility for incorporating additional constraints expressed in a linear form (Blockeel et al. 2023). The recent progress in algorithms designed for solving MILP problems, paired with commercial and academic software packages, has broadened its applicability (Scavuzzo et al. 2024a). Nevertheless, conventional methods often struggle with the computational complexities of large-scale and dynamic real-world scenarios. With its capacity to learn from data, recognize patterns, and make predictions, AI presents complementary tools to improve MILP methodologies. This synergy unlocks new possibilities for optimizing intricate systems, enhancing decision-making processes, and improving operational efficiency (Fan et al. 2024; Scavuzzo et al. 2024a).

The integration of machine learning and mathematical optimization with MILP has spurred substantial interest in both academia and industry, motivated by the potential to enhance the efficiency, scalability, and robustness of optimization systems. The fusion of these fields offers opportunities to develop innovative solution methodologies that leverage the strengths of both machine learning and mathematical optimization (Scavuzzo et al. 2024a). This paper aims to explore the application of AI in solving MILP problems, with a particular emphasis on employing learning to optimize (L2O) techniques with explainability. The objective is to demonstrate the transformation of a MILP instance into a bipartite graph, train structured embeddings of nodes for variables and constraints using Graph Convolutional Networks (GCN) and Graph Attention Network (GAT) and analyze the geometry of the embedding using Instance Space Analysis (ISA).

This paper makes a significant contribution to the field by providing a demonstration of machine learning techniques applied to MILP, focusing on learning methodologies for optimization with explainability. The discussion extends to the application of these techniques in air transport and mobility, presenting a real-world application and demonstrating the practical impact of the discussed approach. The work also analyzes the pipeline as a whole and emphasizes the interactions between the different components (Fan et al. 2024).

Literature Review

*Mixed Integer Linear Programming*

MILP is a foundational mathematical modeling technique for complex decision-making, optimizing linear objectives under linear constraints where some or all variables are restricted to integers, effectively modeling discrete choices (Li et al. 2025). Due to enhanced solver efficiency and accessibility, MILP has become a cornerstone of operations research (Clautiaux and Ljubić 2024).

These models are tackled using a combination of exact methods, such as branch-and-bound (Fan et al. 2024) and cutting planes, and various heuristic algorithms, each presenting distinct strengths and weaknesses (Scavuzzo et al. 2024b, a).

Despite its broad applicability, MILP faces two primary limitations: computational complexity for large-scale instances and the requirement for expert formulation (Li et al. 2024). This drives a continuous demand for more efficient algorithms and solution methodologies, with modern computing architectures like Graphics Processing Unit (GPUs) offering promising avenues for improvement (Clautiaux and Ljubić 2024).

*Artificial Intelligence in Optimization*

AI techniques, particularly Machine Learning (ML), have effectively addressed optimization problems. ML algorithms learn from data to discern patterns (Li et al. 2024), can predict optimal or near-optimal solutions, guide search algorithms, and improve decision-making processes (Martínez-Martínez et al. 2020; Pasupuleti et al. 2024). L2O, a nascent interdisciplinary field bridging classical optimization and machine learning, employs data-trained ML models to augment or supplant hand-engineered algorithms (Chen et al. 2024).

Frequently used AI models in this context include Graph Neural Networks (GNN) and Recurrent Neural Networks (RNN) (Fan et al. 2024). GNNs enables the analysis of spatio-temporal dependencies (Monemi et al. 2025), model connections between transportation entities to optimize networks, identify optimal pathways, predict bottlenecks, and suggest dynamic rerouting (Fan et al. 2024).

Among GNN applications, GCN applies convolution operations directly to graph structures, allowing them to learn node features by aggregating information from local neighborhoods (Li et al. 2024), capturing complex spatiotemporal dependencies and propagation phenomena in dynamic networks, such as flight delay propagation across airports (Wu et al. 2024).

In complement, GAT uses an attention mechanism to weigh the importance of neighboring nodes, focus on relevant connections, like demand prediction and resource allocation (Paul et al. 2024), modeling complex, non-local relationships for accurate origin-destination demand (Smit et al. 2024). For example, an attention mechanism within a sequence-to-sequence framework can capture temporal heterogeneity in network traffic (Bian et al. 2024).

Complementing these learning approaches, ISA is a technique for identifying features that affect solver performance, guiding the development of AI models, and providing AI explainability (xAI). This involves embedding variables and constraint nodes in a latent space to capture local structural patterns and using attention mechanisms over subsampled nodes to efficiently model global dependencies (Li et al. 2024).

*MILP and AI Applications in Air Transportation and Mobility*

Artificial intelligence is essential for optimizing air transportation and mobility systems. Airlines use AI algorithms to analyze large volumes of data, such as flight histories and weather patterns, to create optimized schedules that minimize delays, maximize aircraft utilization, and enhance profitability.

Concurrently, within the complex ecosystems of airports, AI algorithms process real-time data regarding flights and passengers to optimize resource allocation, including gates and personnel. This combined approach results in an improved passenger experience, reduced delays, and an overall increase in operational efficiency across the sector (Geske et al. 2024).

For the Crew Pairing Problem (CPP), Pereira et al. (2022) address the trade-off between fast, low-quality solvers and slow, high-quality ones. A ML model was used to imitate the decisions of a slow but expert optimization method, cutting CPU time by approximately 74%.

Following the pairing stage, for the Crew Rostering Problem (CRP), Racette et al. (2025) use a windowing technique. First, a ML model generates an initial complete roster. This roster then guides a branch-and-price algorithm that re-optimizes the schedule in smaller, overlapping time windows. The initial solution provides valuable information, enabling the combined approach to be over 10 times faster than the state-of-the-art solver while delivering solutions on average less than 1% from an optimal result.

*Gaps in the Literature*

Although existing literature highlights the potential of integrating AI with MILP, several research gaps remain to be addressed. Further research is warranted to explore the application of AI to model formulation and enhancement for specific algorithms, such as the Alternating Direction Method of Multipliers and column generation (Fan et al. 2024). Furthermore, it is crucial to address ethical considerations, including user information protection, and to integrate informed consent and human autonomy into community engagement programs to foster sustainability in this sector (Mirindi 2024). The gap is evident as current research often concentrates on enhancing specific algorithms or high-level applications instead of creating integrated frameworks suitable for systems with critical safety concerns (Singh 2024).

Consequently, practical challenges like data dependence, model interpretability and explainability, and transparent decision-making in dynamic environments remain insufficiently addressed (Singh 2024; Mirindi 2024). This emphasizes the need to go beyond isolated improvements, advocating for robust and scalable AI algorithms that integrate adaptive learning with the formal assurances of mathematical optimization, particularly given the complexity and uncertainty inherent in real-world optimization problems (Singh 2024; Mirindi 2024).

Aiming to fill this gap, this article adapts the Instance Space Analysis (ISA) framework to extract and visualize the structural complexity of a specific air problem. This serves to demonstrate the potential of xAI as an initial step in applying learning to optimize (L2O) for MILP solving.

## Methodology

We outline the computational process created to convert a MILP instance into a low-dimensional embedding space that is appropriate for ISA using the air05 instance from the Mixed Integer Programming Library (MIPLIB)[1]. The process consists of three main steps: (1) modeling the MILP as a heterogeneous bipartite graph, (2) creating high-dimensional node embeddings with GNNs, and (3) mapping these embeddings into a 2D space for visualization purposes.

*MILP Instance as a Bipartite Graph*

The approach begins with the air05 instance, which is accessed with a Gurobi Academic License from an MPS file, and addresses a set partitioning problem related to scheduling airline crews. It serves as a benchmark for evaluating the performance of various mixed integer programming solvers. This instance is based on real-world scenarios where crews need to be assigned to multiple flights, ensuring that all flights are adequately staffed while adhering to operational and regulatory requirements, such as work hours and rest periods. The main goal is to minimize the overall cost of these assignments.

We first represent this optimization problem as a heterogeneous bipartite graph to illustrate the connections between variables and constraints. To construct and manage this data structure, we leverage the PyTorch Geometric (PyG) library[2], which is designed to handle heterogeneous graphs. The nodes are divided into two types: variable nodes, which represent each decision variable, and constraint nodes, which correspond to each linear constraint. The edges indicate the non-zero entries in the constraint matrix A. This setup creates two types of directed edges for message passing: an edge of type ("var", "inc", "con") exists when a variable is involved in a constraint, and a corresponding reverse edge of type ("con", "rev_inc", "var") is also established.

---

[1] https://miplib.zib.de/instance_details_air05.html

Initial features are derived directly from the MILP formulation to supply the GNN with the necessary input data. For each variable node, we create a feature vector that includes its objective coefficient, lower bound (LB), upper bound (UB), and an integrality flag (e.g., 1.0 for integer/binary variables and 0.0 for continuous ones). For each constraint node, the feature vector consists of its right-hand-side (RHS) value and its sense, which is numerically encoded (for instance, "less than or equal to" as -1, "equal to" as 0, and "greater than or equal to" as 1). The non-zero coefficient from the constraint matrix is used as a scalar edge feature. To maintain numerical stability and control the magnitudes of these features while preserving sign information, all edge weights are processed through a hyperbolic tangent (tanh) activation function.

*Generating Embeddings with GNNs*

We represent instances as graphs and utilize a GNN as a feature extractor that is aware of the underlying structure. It's important to clarify that in this study the GNN is not yet employed to solve the optimization problem like in L2O. Instead, it focuses on learning high-dimensional embedding vectors for all nodes, which reflect their topological relationships and similarities within the problem's structure. We implement and evaluate two different heterogeneous GNN encoder architectures, built upon the bipartite graphs in PyG:

- GCN: This model, which we refer to as "HeteroGraphConv" in our implementation, stacks standard "GraphConv" layers adapted for bipartite message passing between variable and constraint node types.
- GAT: This model, referred to as "HeteroGAT", employs "GATConv" layers. These layers incorporate an attention mechanism, enabling nodes to assign varying weights to messages from different neighbors, potentially capturing more subtle relationships.

---

[2] https://www.pyg.org/

To train these GNNs for representation, we use a self-supervised link prediction task. We take the true edges (var, inc, con) from the MILP as positive samples and generate an equal number of negative samples by randomly selecting (var, con) pairs that do not exist in the sparse constraint matrix. The GNNs are trained to create final node embeddings ($Z_{var}$, $Z_{con}$) in such a way that the dot-product similarity of connected (positive) pairs is maximized, while the similarity of unconnected (negative) pairs is minimized. This is achieved using a Binary Cross-Entropy with Logits (BCE with logits) loss function. The training process ensures that the final embeddings capture the connectivity structure of the instance.

*Instance Space Analysis*

ISA is a methodology used to relate instance features to algorithm performance to understand the strengths and weaknesses of different algorithms for explainability (Liu et al. 2024; Christiansen and Smith-Miles 2025). Inspired by its techniques, we adapt the visualization techniques from ISA to analyze the high-dimensional embedding matrices $Z_{var}$ and $Z_{con}$, which are generated by the trained GNNs.

Since these vectors are not easily interpretable, we utilize dimensionality reduction methods to map them into a two-dimensional space for visual examination. We employ three main techniques to offer different perspectives on the data:

- Principal Component Analysis (PCA): This linear method helps identify the directions of maximum variance in the data. It highlights the main linear structure and overall distribution of the embedding space.
- Uniform Manifold Approximation and Projection (UMAP): A non-linear manifold learning technique used to preserve both the local neighborhood relationships and the global, large-scale structure of the data.
- t-Distributed Stochastic Neighbor Embedding (t-SNE): Also a non-linear technique for manifold learning that aims to maintain the local neighborhood relationships. It uncovers distinct clusters, or "islands," of nodes that are closely related in the high-dimensional space, even if they cannot be separated linearly.

The resulting two-dimensional (2D) scatter plots form what we refer to as the "Instance Space", albeit here it still does not include performance features. In this space, each point corresponds to either a single variable or a constraint, with its location reflecting its learned structural role within the MILP instance.

To assess the congruence between the derived geometric representations and predefined discrete structural roles, a visualization approach involving the coloring of points based on automatically generated clusters was employed, namely k-means. Prior to this, the following steps were performed to identify the best data clustering:

- The Elbow Method determines an appropriate number of clusters for the $Z_{v}arg_{cn}$ embeddings. This approach involves plotting the Within-Cluster Sum of Squares (WCSS) against the number of clusters. WCSS quantifies the compactness of clusters. As the number of clusters $k$ increases, WCSS typically decreases. The "elbow point," where the rate of WCSS decrease significantly diminishes, often indicates the optimal *k*.

- Following this, the Silhouette Method is employed to determine the optimal number of clusters for the $Z_var_gcn$ embeddings. This method computes a silhouette score for each data point, indicating its similarity to its own cluster relative to other clusters. Scores range from -1 to 1, with higher values signifying superior clustering. The optimal is typically identified as the value yielding the highest average silhouette score across different *k* values.

This two-step clustering validation ensures that the identified structural roles within the MILP instances are both compact and well-separated, establishing a foundation for subsequent analysis and AI model development (Gupta et al. 2022).

## Results

*Instance Profile and Graph Characteristics*

The air05 instance includes 7,195 integer variables and 426 constraints, along with 52,121 non-zero elements in its constraint matrix. Figure 1 shows a visualization of the air05 instance as a bipartite graph, with constraint nodes (red) and variable nodes (blue).

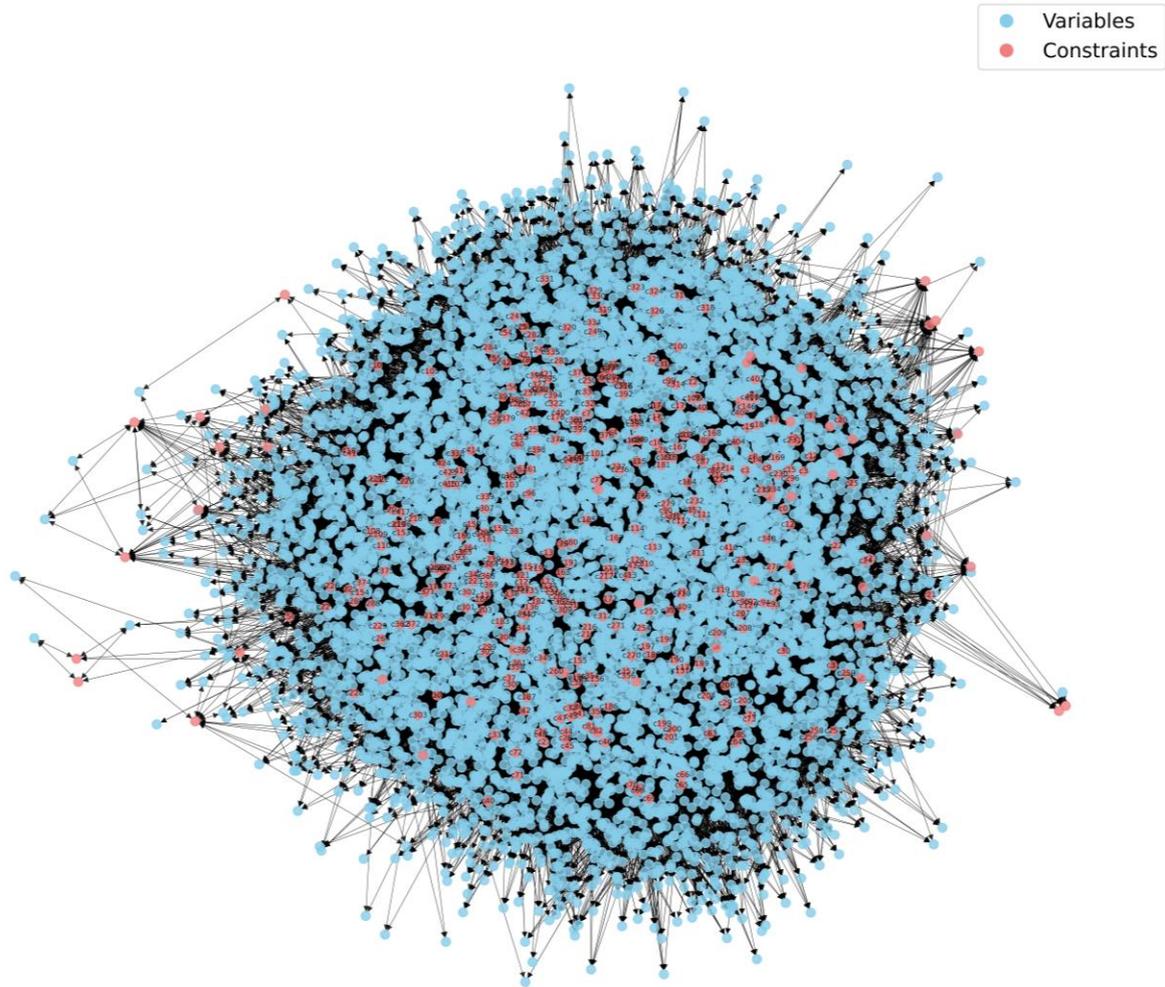

Source: Elaborated by the authors (2025).

**Figure 1**. Bipartite graph.

The edges represent the connections between variables and the constraints they appear in. This layout illustrates the problem's dense and complex connectivity. Figure 2 shows the sparsity plot of the bipartite adjacency matrix. In this plot, the Y-axis displays the constraints, while the X-axis shows variables. Each dot represents a non-zero entry ($A_{ij}$) in the constraint matrix, which corresponds to an edge in Figure 1.

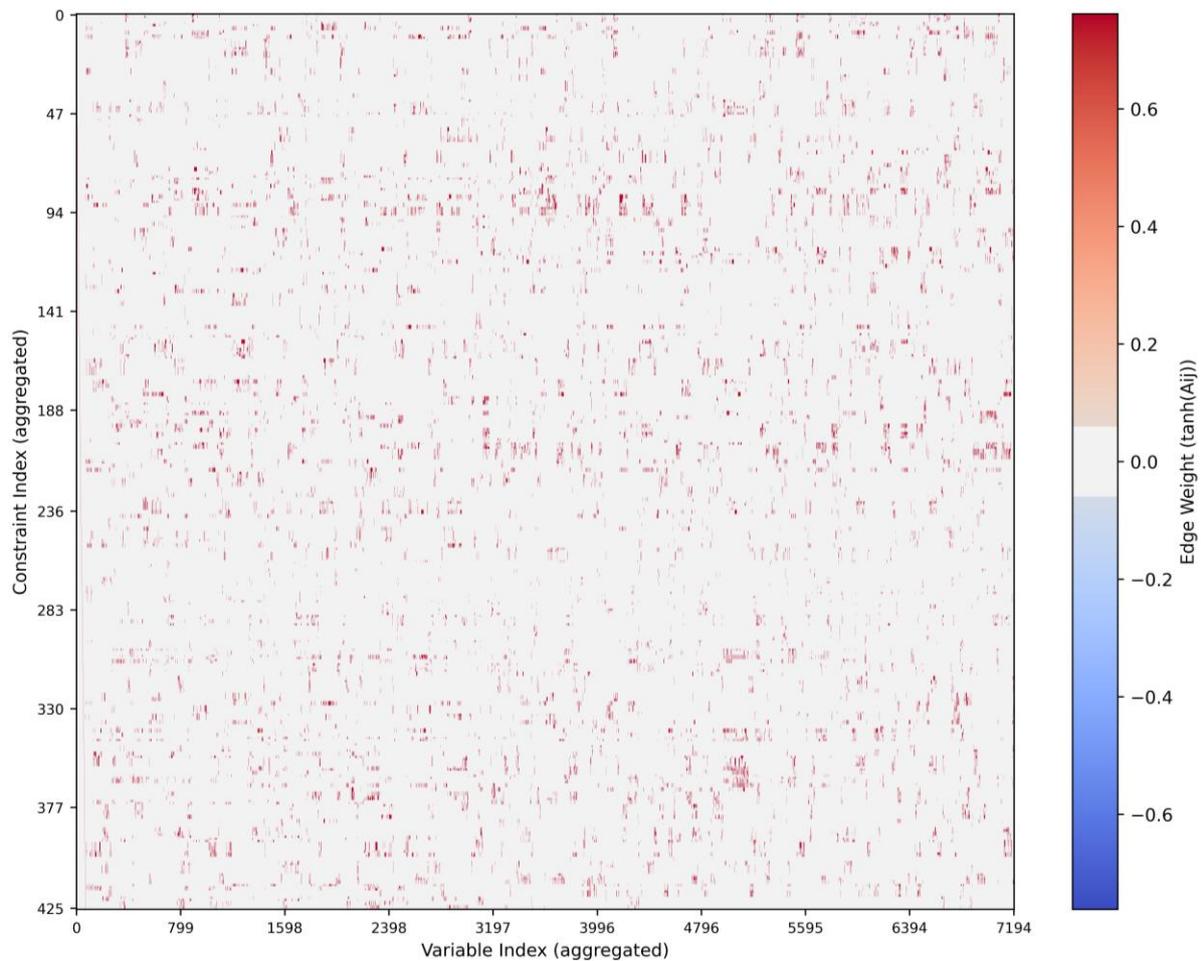

Source: Elaborated by the authors (2025).

**Figure 2**. Bipartite adjacency matrix plot: constraints and variables.

Two important characteristics of the instance are evident. First, the graph is very sparse, with the 52,121 non-zero entries making up only 1.70% of the nearly 3.065 million potential entries in the matrix. This level of sparsity is expected for large-scale combinatorial optimization problems. Second, the non-zero entries are not evenly distributed. It is possible to clearly distinguish vertical lines (dense columns) that represent specific variables involved in many constraints. This uneven and complex structure highlights the intricacy of the problem and supports the decision to use a graph-based learning method, as GNNs can manage such complex and irregular topologies.

*GNN Model Training and Validation*

Before analyzing the resulting embeddings, it is necessary to validate that the GNN models successfully learned from the graph structure. The training loss for the link prediction pretext task is presented in Fig. 3.

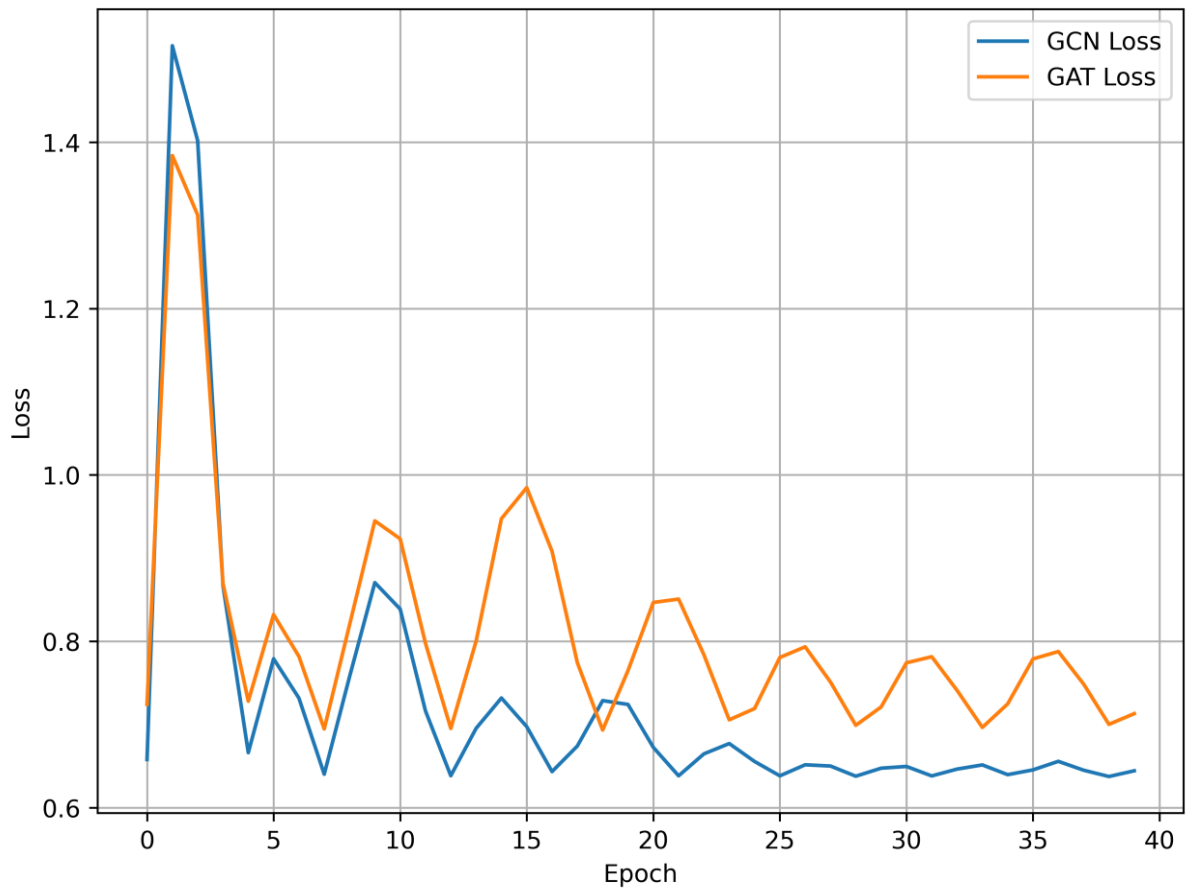

Source: Elaborated by the authors (2025).

**Figure 3**. Training loss.

Figure 3 plots the training loss curves (binary cross-entropy for link prediction) for the GCN and GAT models. Both architectures display a similar learning pattern. There is a noticeable and rapid reduction in loss during the first five epochs, showing that both models quickly recognized and started to fix major errors in their link predictions. After this initial drop, the loss continues to decrease, but with significant fluctuations, especially in the GAT model, which shows higher amplitude oscillations compared to the GCN model. This variability indicates that the GAT model might be dealing with a more complicated or unstable optimization process for this task.

Despite the fluctuations, both models show a clear trend of convergence, as the loss steadily decreases and starts to stabilize in the later epochs. The GCN model achieves a lower and more stable loss compared to the GAT model. This stabilization suggests that the models have moved beyond random guessing and have successfully learned patterns from the graph's connections. The effective convergence of both models on this self-supervised pretext task confirms that their resulting node embeddings have captured the important structural and topological features of the air05 instance, making them suitable for further analysis.

*Analysis of the Learned Instance Space*

After completing the model training validation, the ISA starts by visualizing the high-dimensional node embeddings. The embeddings for both variable (VAR) and constraint (CON) nodes, produced by the GCN and GAT models, were initially grouped using a k-means algorithm. Figure 4 shows the first visualization of these embeddings, utilizing Principal Component Analysis (PCA) for linear dimensionality reduction. The resulting 2D projections are color-coded based on their designated k-means cluster (k=10).

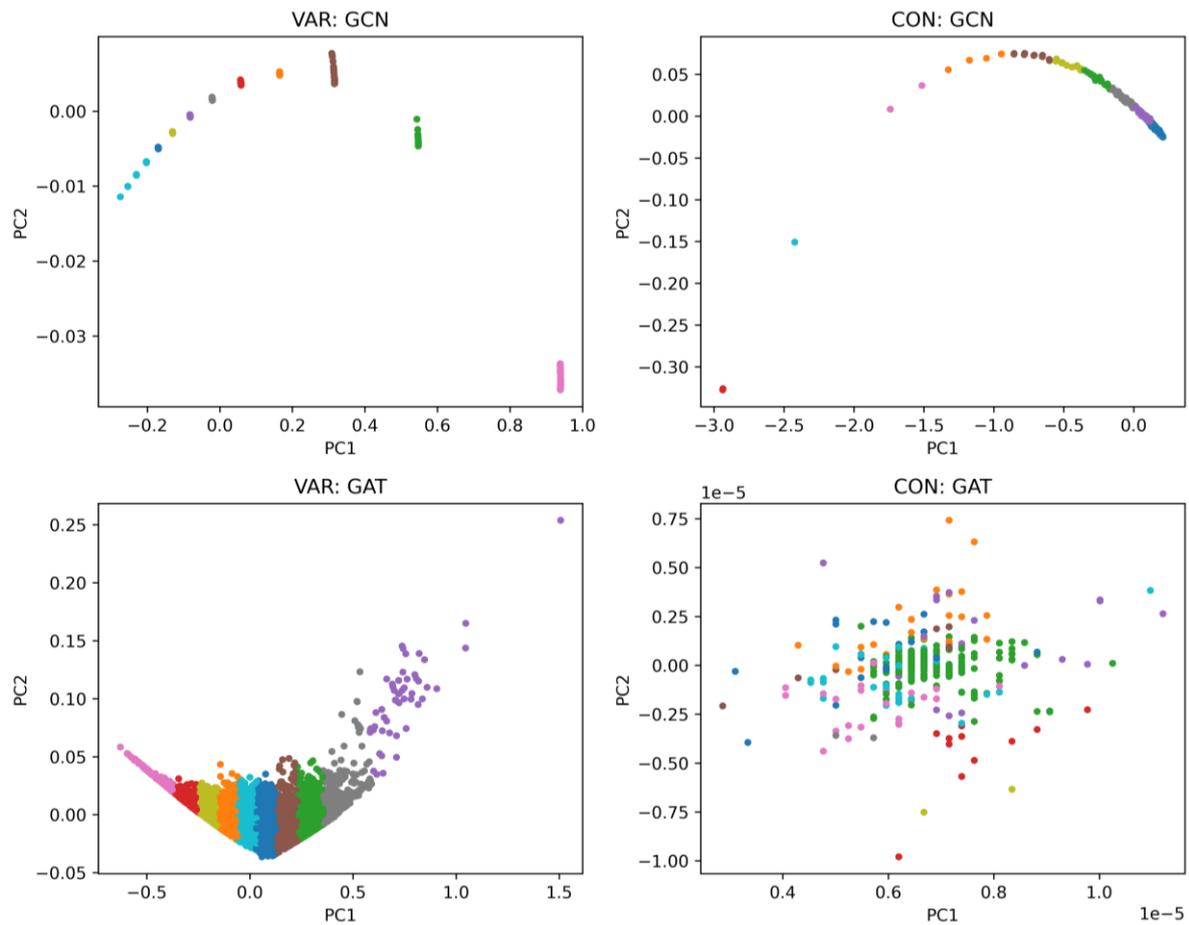

Source: Elaborated by the authors (2025).

**Figure 4**. 2D projections of VAR and CON node embeddings from the GCN and GAT models, reduced using PCA and clustered using k-means.

The linear projections of the GCN model, illustrated in the top row of Fig. 4, display the initial structure identified by the principal components. In the "VAR: GCN" plot, the variable embeddings take on several distinct, arc-like formations. However, the clusters are closely stacked along these arcs, showing significant overlap and limited linear separability. This issue of separation is similarly reflected in the "CON: GCN" plot, where the constraint embeddings also align into a clear arc but do not separate the clusters, as they are compressed along the same structure.

The results from the GAT model, displayed in the bottom row, show a significant amount of overlap among the clusters. In the "VAR: GAT" plot, nearly all variable node clusters merge into a compact, fan-shaped formation, making it very difficult to differentiate between the various colored groups. This problem is even more evident in the "CON: GAT" plot, where the constraint embeddings create a chaotic and disordered cloud. In this plot, all clusters blend into a single intertwined mass, revealing almost no distinct structure.

The overlap seen suggests that the structural roles identified by the k-means algorithm can't be separated in a straightforward way in the original high-dimensional space. While the PCA projection does a good job of capturing the main variance axes (PC1 and PC2), it fails to clearly differentiate the various cluster structures. This indicates that the GNNs have created a complex, non-linear representation of the relationships among nodes. Consequently, this finding highlights the necessity for non-linear dimensionality reduction techniques such as UMAP and t-SNE, which are designed to uncover these complex geometries while preserving the local neighborhood structures.

Given the limitations of linear projection revealed by PCA, we next apply UMAP, a non-linear technique designed to preserve the topological and local structure of the data, as shown in Fig. 5.

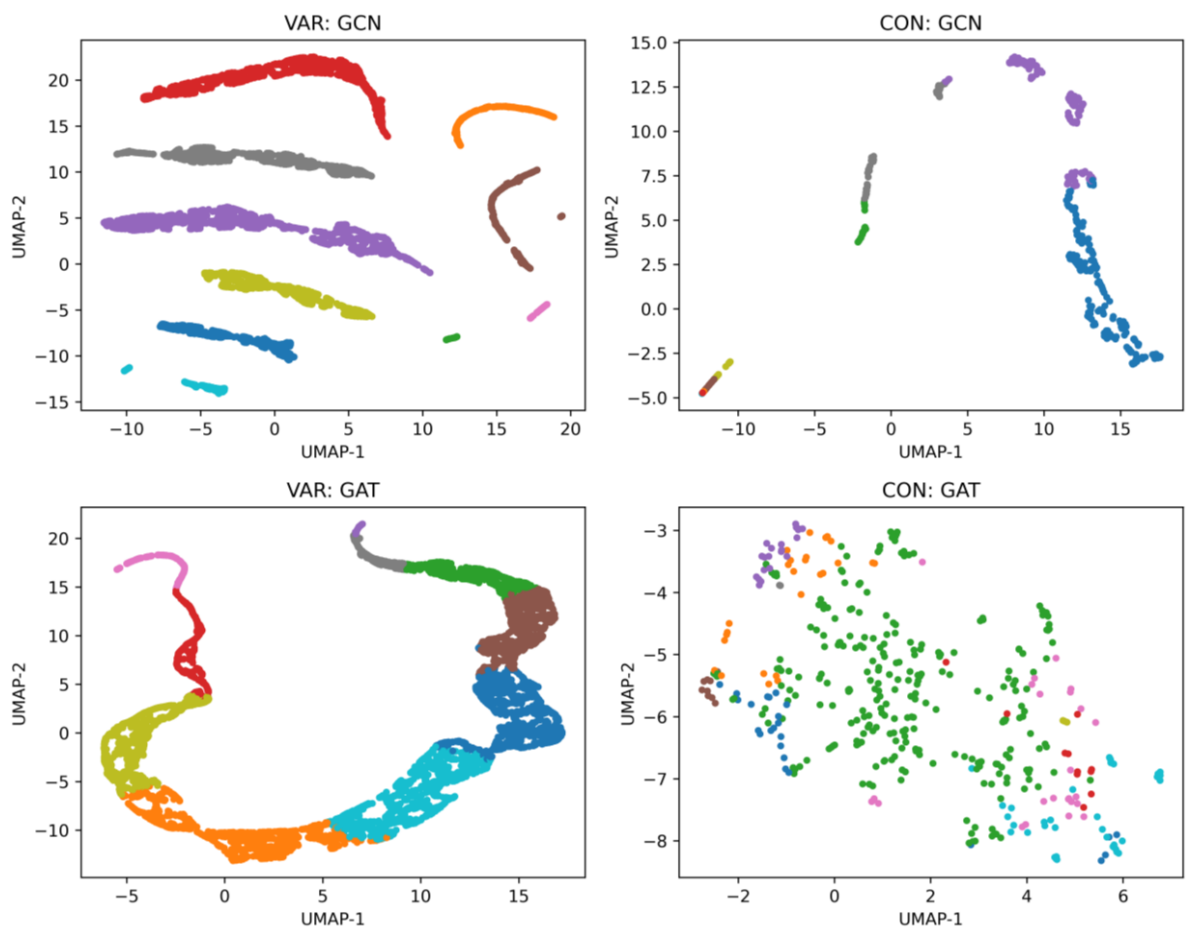

Source: Elaborated by the authors (2025).

**Figure 5**. 2D projections of VAR and CON node embeddings from the GCN and GAT models, reduced using UMAP and clustered using k-means.

Figure 5 complements the PCA analysis by applying UMAP (Uniform Manifold Approximation and Projection). This non-linear method reveals the local neighborhood structures that were not visible in the linear projections.

The results from the GCN model, illustrated in the top row of Fig. 5, reveal a well-organized embedding space. In the "VAR: GCN" plot, the variable embeddings are displayed as several clearly defined "islands," with each colored island representing one of the 10 k-means clusters. This distinct visual separation suggests that the GCN model effectively captured a complex, non-linear structure, allowing for a clear distinction between different functional groups of variables. Similarly, the "CON: GCN" plot shows the constraint node embeddings, which also display well-defined and separate clusters. This indicates that the GCN model was successful in distinguishing between the various types of constraints present in the problem.

The embeddings from the GAT model, shown in the bottom row, reveal a more intricate scenario. The "VAR: GAT" plot effectively distinguishes the clusters, indicating that the model has learned a separable structure for variable nodes. However, the geometry observed is quite different from that of the GCN, with the clusters forming a long, continuous "S"-shaped pattern. This difference in geometry is especially noticeable in the "CON: GAT" plot. Unlike the other three subplots, the embeddings for the constraints in the GAT model appear much less organized. The clusters are dispersed and significantly overlap, indicating that this model struggled to learn a clear, high-level representation for the constraint nodes.

This visualization highlights that GNN models captured a complex, non-linear geometry, although the GAT model's success was limited to the variable nodes, as its constraint embeddings remained disorganized. The closeness of the clusters is significant, suggesting that adjacent clusters correspond to node roles that are functionally similar, while isolated clusters indicate more distinct structural roles.

To provide an alternative non-linear viewpoint and further confirm the coherence of the k-means cluster assignments, we also utilized the t-SNE algorithm, with the results displayed in Fig. 6.

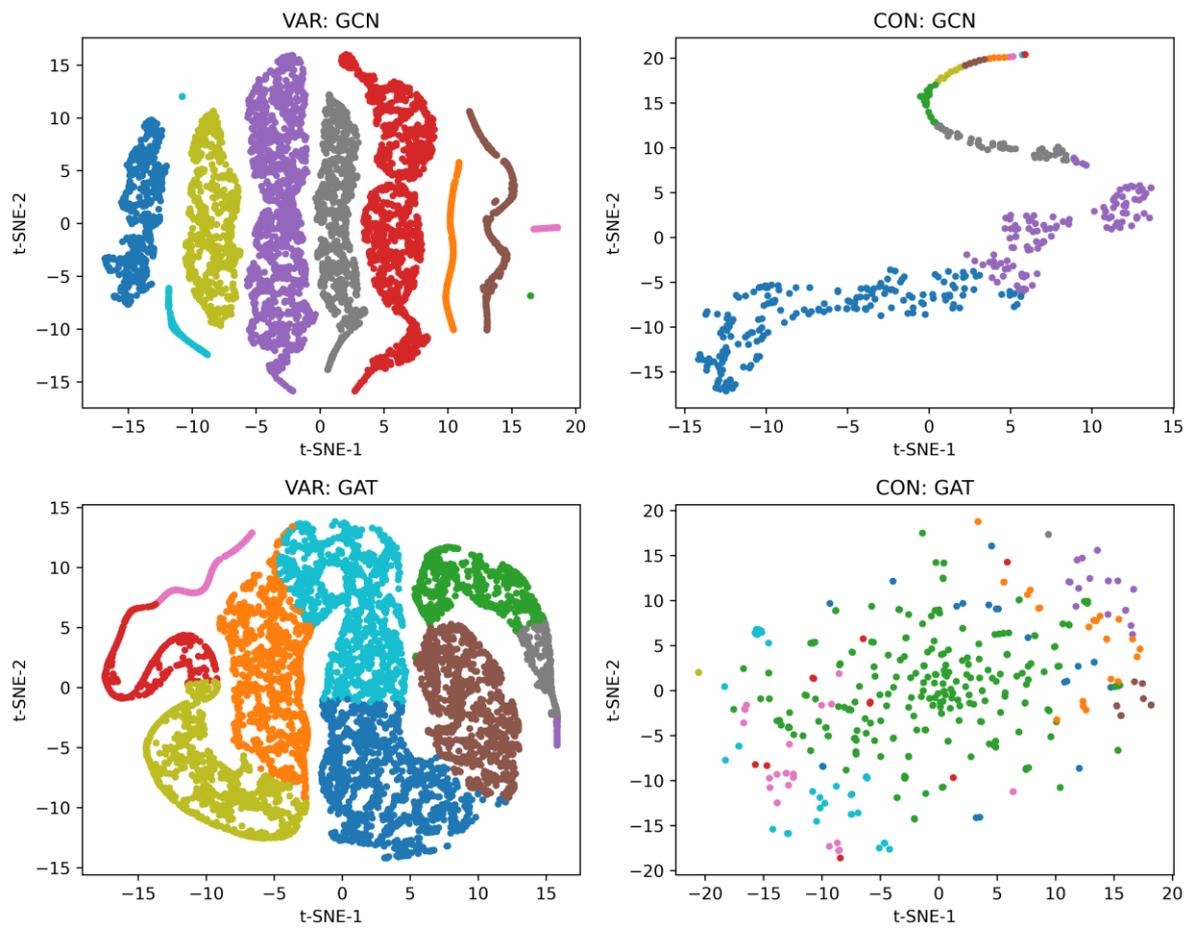

Source: Elaborated by the authors (2025).

**Figure 6**. 2D projections of VAR and CON node embeddings from the GCN and GAT models, reduced using t-SNE and clustered using k-means.

Figure 6 presents a third visualization using t-SNE (t-distributed Stochastic Neighbor Embedding). Similar to UMAP, t-SNE is a non-linear technique that emphasizes local structure; however, its objective function differs, often resulting in tighter, more distinctly separated cluster visuals.

The t-SNE projections for the GCN model, shown in the top row, strongly validate the findings from UMAP. The "VAR: GCN" plot again reveals the clusters as dense, compact, and well-defined "islands" that are clearly separated from one another. This reinforces that the GCN model learned a strong, separable structure for the variable nodes. The "CON: GCN" plot also shows clear separation, with the constraint nodes organized into a few distinct groups that appear more defined than in the UMAP visualization, validating the coherence of the constraint groupings.

The GAT model's results, in the bottom row, also mirror the UMAP analysis. The "VAR: GAT" plot successfully visualizes the clusters as distinct, dense-bodied groups, confirming that a separable structure was learned for the variable nodes (albeit with a different geometry than the GCN). However, the "CON: GAT" plot provides the same stark contrast seen in the previous non-linear projection. It shows a highly disorganized and chaotic cloud of points, with all 10 k-means clusters completely intermingled. This lack of structure strongly confirms that the GAT model failed to learn a separable representation for the constraint nodes.

## Discussion

The results of this study present three primary implications for the application of AI in air transportation optimization.

First, the methodology validates the use of bipartite graph embeddings as a viable alternative to manual feature engineering. Feature extraction processes in previous studies required domain-specific descriptors to map algorithm footprints (Liu et al. 2024). The mapping of the air05 instance, consistent with findings on benchmarks from MIPLIB 2017, demonstrates that GNNs can automatically extract structural features from raw MILP formulations (Deza and Khalil 2023). This aligns with established frameworks where instance feature vectors enable the visualization of problem spaces and the assessment of benchmark diversity (Smith-Miles and Muñoz 2023). The approach reduces the reliance on domain-specific heuristics, offering a scalable pathway to analyze large sets of heterogeneous instances. However, recent work suggests that some raw formulations may require feature augmentation, such as unique identifiers or coloring schemes, to fully capture structural expressiveness (Han et al. 2025).

Second, the performance difference between the two architectures suggests that simpler models may be more effective for this specific class of sparse bipartite graphs. The GCN model effectively captured the global topology—variables and constraints—whereas the GAT model failed to organize the constraint space. This indicates that for initial structural encoding of MILP instances, attention mechanisms may introduce unnecessary complexity without yielding better representations for explainability.

Finally, the distinct clustering observed in the ISA confirms that the learned embeddings capture the functional roles of variables and constraints, significantly contributing to xAI. Previous applications of ISA indicate that identifying such structural properties allows for the prediction of algorithm performance across different regions of the instance space (Christiansen and Smith-Miles 2025). This structural awareness is a prerequisite for L2O applications. Aggregating variables into higher-level decision structures has been shown to enhance branching quality in vehicle routing problems (Silva et al. 2025). These embeddings can serve as the foundation for training agents to make informed decisions, such as variable selection for branching or generating warm-start solutions, directly addressing the computational challenges of real-world solvers.

## Conclusion

The study assessed the feasibility of integrating AI with MILP to address optimization challenges in air transportation with explainability. By transforming the air05 crew scheduling instance into a heterogeneous bipartite graph, the research demonstrated that GNNs can extract structural features, offering an alternative to manual feature engineering. The methodological analysis indicated distinct performance characteristics between the tested architectures. While the GCN effectively captured the global topology, producing well-defined and separable clusters for both variables and constraints, the GAT struggled to organize the constraint space, suggesting that simpler models may be more effective for this specific class of sparse bipartite graphs.

Regarding visualization, PCA identified the main variance axes but was unable to distinguish specific cluster structures, indicating the complexity of the embedding space. Consequently, applying non-linear reductions like UMAP and t-SNE proved essential to effectively visualize the high-dimensional embeddings, confirming that the learned representations reflect the functional roles of decision variables.

Beyond technical feasibility, these findings hold important implications for the operational efficiency and sustainability of the air mobility sector. The ability to automate feature extraction and improve solver behavior directly contributes to optimizing complex tasks, such as crew scheduling and airport resource allocation, which are critical for maximizing profitability and minimizing delays.

Future research should focus on developing explainable L2O agents that utilize these structural embeddings for specific solver tasks, such as variable selection for branching or generating warm-start solutions, thereby directly addressing computational challenges. This development could be supported by investigating feature augmentation techniques, such as unique identifiers, to enhance structural expressiveness, while expanding analysis to a broader diversity of heterogeneous instances to test generalization capabilities.

References

Bian Z, Gao J, Ozbay K, Li Z (2024) Traffic Prediction considering Multiple Levels of Spatial-temporal Information: A Multi-scale Graph Wavelet-based Approach

Blockeel H, Devos L, Frénay B, et al (2023) Decision trees: from efficient prediction to responsible AI. Front Artif Intell 6:1124553. https://doi.org/10.3389/frai.2023.1124553

Chen X, Liu J, Yin W (2024) Learning to optimize: A tutorial for continuous and mixed-integer optimization

Christiansen J, Smith-Miles K (2025) Instance Space Analysis for the Quadratic Assignment Problem

Clautiaux F, Ljubić I (2024) Last fifty years of integer linear programming: A focus on recent practical advances. European Journal of Operational Research 324:707–731. https://doi.org/10.1016/j.ejor.2024.11.018

Deza A, Khalil EB (2023) Machine Learning for Cutting Planes in Integer Programming: A Survey. In: Proceedings of the Thirty-Second International Joint Conference on Artificial Intelligence. International Joint Conferences on Artificial Intelligence Organization, Macau, SAR China, pp 6592–6600

Fan Z, Ghaddar B, Wang X, et al (2024) Artificial Intelligence for Operations Research: Revolutionizing the Operations Research Process

Geske AM, Herold DM, Kummer S (2024) Artificial intelligence as a driver of efficiency in air passenger transport: A systematic literature review and future research avenues. Journal of the Air Transport Research Society 3:100030. https://doi.org/10.1016/j.jatrs.2024.100030

Gupta S, Modgil S, Bhattacharyya S, Bose I (2022) Artificial intelligence for decision support systems in the field of operations research: review and future scope of research. Ann Oper Res 308:215–274. https://doi.org/10.1007/s10479-020-03856-6

Han Q, Li Q, Yang L, et al (2025) Feature Augmentation of GNNs for ILPs: Local Uniqueness Suffices

Li H, Zhao Y, Mao Z, et al (2024) Graph Neural Networks in Intelligent Transportation Systems: Advances, Applications and Trends

Li S, Kulkarni J, Menache I, et al (2025) Towards Foundation Models for Mixed Integer Linear Programming

Liu C, Smith-Miles K, Wauters T, Costa AM (2024) Instance space analysis for 2D bin packing mathematical models. European Journal of Operational Research 315:484–498. https://doi.org/10.1016/j.ejor.2023.12.008

Martínez-Martínez V, Nevares I, Del Alamo-Sanza M (2020) Artificial Intelligence Methods for Constructing Wine Barrels with a Controlled Oxygen Transmission Rate. Molecules 25:3312. https://doi.org/10.3390/molecules25143312

Mirindi D (2024) A Review of the Advances in Artificial Intelligence in Transportation System Development. JCCEE 9:72–83. https://doi.org/10.11648/j.jccee.20240903.13

Monemi RN, Gelareh S, González PH, et al (2025) Graph Convolutional Networks for logistics optimization: A survey of scheduling and operational applications. Transportation Research Part E: Logistics and Transportation Review 197:104083. https://doi.org/10.1016/j.tre.2025.104083

Pasupuleti V, Thuraka B, Kodete CS, Malisetty S (2024) Enhancing Supply Chain Agility and Sustainability through Machine Learning: Optimization Techniques for Logistics and Inventory Management. Logistics 8:73. https://doi.org/10.3390/logistics8030073

Paul S, Witter J, Chowdhury S (2024) Graph Learning-based Fleet Scheduling for Urban Air Mobility under Operational Constraints, Varying Demand & Uncertainties

Pereira P, Courtade E, Aloise D, et al (2022) Learning to branch for the crew pairing problem. Les Cahiers du GERAD

Racette P, Quesnel F, Lodi A, Soumis F (2025) Accelerated windowing for the crew rostering problem with machine learning

Scavuzzo L, Aardal K, Lodi A, Yorke-Smith N (2024a) Machine learning augmented branch and bound for mixed integer linear programming. Math Program. https://doi.org/10.1007/s10107-024-02130-y

Scavuzzo L, Aardal K, Yorke-Smith N (2024b) Learning optimal objective values for MILP

Silva JMP, Uchoa E, Subramanian A (2025) Cluster Branching for Vehicle Routing Problems. INFORMS Journal on Computing ijoc.2024.1036. https://doi.org/10.1287/ijoc.2024.1036

Singh R (2024) Artificial Intelligence In Operations Research: Bridging The Gap Between Theory And Practice. Multidisciplinary International Journal 10:73–81

Smit IG, Zhou J, Reijnen R, et al (2024) Graph Neural Networks for Job Shop Scheduling Problems: A Survey

Smith-Miles K, Muñoz MA (2023) Instance Space Analysis for Algorithm Testing: Methodology and Software Tools. ACM Comput Surv 55:1–31. https://doi.org/10.1145/3572895

Wu Y, Yang H, Lin Y, Liu H (2024) Spatiotemporal Propagation Learning for Network-Wide Flight Delay Prediction. IEEE Transactions on Knowledge and Data Engineering 36:386–400. https://doi.org/10.1109/TKDE.2023.3286690